\documentclass[aps,prl,twocolumn,float]{revtex4}
\usepackage{amsmath,bm,epsfig}
\newcommand{\B}[1]{{\bm{#1}}}
\usepackage[dvips]{color}
\begin{document}

\title{Mechanical Yield in Amorphous Solids: a First-Order Phase Transition}
\author{Prabhat Jaiswal, Itamar Procaccia, Corrado Rainone and Murari Singh}
\affiliation{Dept. of Chemical Physics, The Weizmann Institute of Science,  Rehovot 76100, Israel}
\begin{abstract}
Amorphous solids yield at a critical value of the strain (in strain controlled experiments); for
larger strains the average stress can no longer increase - the system displays an elasto-plastic steady state.
A long standing riddle in the materials community is what is the difference between the microscopic states
of the material before and after yield.
Explanations in the literature are material specific, but the universality of the phenomenon begs
a universal answer.
We argue here that there is no fundamental difference in the states of matter before and after yield, but the yield is a {\em bona-fide} first order phase transition between a highly restricted set of possible configurations residing in a small region of phase space to a vastly rich set of configurations which include many marginally stable ones. To show this we employ an order parameter of universal applicability, independent of the microscopic interactions,  that is successful in quantifying the transition in an unambiguous manner.
\end{abstract}
\maketitle
\begin{figure}
\vskip -0.5 cm
\includegraphics[scale = 0.50]{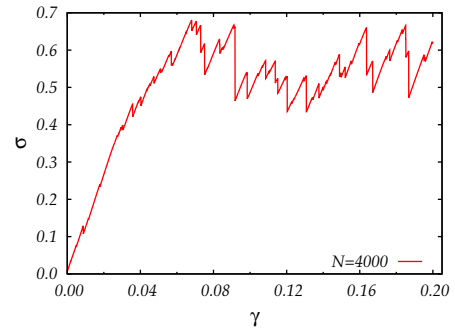}
\caption{A typical stress vs. strain curve obtained in a strain-controlled athermal quasistatic (AQS)
shearing protocol using a Kob-Andersen 65-35\% Lennard Jones Binary Mixture of 4000 particles in $2d$. Note the generic existence of a pre-yield branch in which the stress is increasing on the average when the strain increases, and
a post yield steady state where the average stress is constant. Both regions are punctuated by plastic
events, with the stress drops being much larger in the post-yield compared to the pre-yield branch. This kind
of stress vs. strain curve is ubiquitous for a huge variety of amorphous solids. Here and in the text
 we drop tensorial indices from the stress and the strain for notational simplicity.}
\label{stress-strain}
\end{figure}

A ubiquitous, and in fact universal, characteristic of the mechanical properties of amorphous solids is their
stress  vs. strain dependence \cite{98Ale}. Measured in countless quasi-static strain-controlled simulations (see for example \cite{04VBB,04ML,05DA,06TLB,06ML,09LP,11RTV}) and experiments (see for example \cite{06SLG,13KTG,13NSSMM}),
it typically exhibits two distinct regions. In one region, at lower strain values, the stress $\B \sigma$ increases on the average upon the increase of strain $\B \gamma$, although this increase is punctuated by plastic events. A second region, at higher values of the strain,
displays a constant (on the average) stress which cannot increase even though the strain keeps increasing. Of course
also this elasto-plastic steady state branch is punctuated by plastic events. A typical such shear stress vs. shear strain curve at zero temperature is shown in Fig.~\ref{stress-strain}. The two regions are separated by what is referred to as ``yield". The actual shape of the stress vs. strain curve near the yield point depends on details of the system preparation. Amorphous solids
prepared by a slow quench from the melt tend to display a stress peak before yielding, whereas those prepared
by a fast quench join the steady state smoothly without a stress peak \cite{13JBP}. Of course the steady state branch itself
is independent of the preparation protocol; memory of the initial state is lost in this regime.

The phenomenon of the mechanical yield in
amorphous solids has been a subject of extensive study in recent years. Many
numerical studies on the subject have been performed using athermal,
quasi-static shear (AQS) protocols, wherein a glass is made by quenching a glass
former down to zero temperature, and then subjecting it to a quasi-static
($\dot\gamma \to 0$) shear protocol wherein the system is subject to small shear
steps, and then allowed, after each step, to find a new mechanically stable
minimum of its potential energy. This kind of protocol always gives rise to the same basic kind of
phenomenology as seen in Fig.~\ref{stress-strain} independently of the detailed microscopic
interaction between the constituents.
This basic phenomenology has been reported in very many publications, and there
is a general consensus about the fact that a \emph{qualitative} change of some
sort must take place between the ``elastic" and ``steady-state" parts of the
stress-strain curve. As much as the qualitative picture is evident, however, a
lot of difficulty arises when trying to capture this qualitative change in a
quantitative manner.

Devising a way to distinguish and study two different states of matter, and the
transition that connects one with the other, means identifying an \emph{order
parameter} to act as a label for the states.
This, however is a challenging program in the present case: as much as the ``elastic" and ``steady
state" branches look different (one is able to increase the stress under a shear load, while the other
cannot), a snapshot, say, of a particle configurations in both regimes is unable
to detect any relevant difference between the two. Since both states are anyway
amorphous, there is no trivial order parameter that would allow us to
unambiguously tell them apart. Notice how this difficulty is not present in the
case of crystalline solids, which \emph{do} exhibit evident structural
peculiarities with respect to liquids, and whose mechanism of failure is well
known since decades. Ultimately, it all boils down to finding a suitable order
parameter for the glass phase.

The problem of finding an order parameter for a glass is both practical and
conceptual. First, as we said before, a snapshot of a typical glass
configuration before and after yield does not show
any difference: all glasses look the same to us, and
all of them look like a liquid. Standard methods like structure functions, higher
order correlation functions, Voronoi tesselations, Delaunay triangulations etc. all
failed to provide distinction
between typical configurations before and after yield.

In this Letter we propose that the difficulty in making a distinction between the
pre- and post-yield configurations lies in the fact that {\em there is really no distinction}.
The crux of the matter is not in the nature of configurations but in their number. The yield
takes place because of a sudden opening up of a vast number of configurations that are not
at the system's disposal before yield. To establish this insight we employ an order
parameter of a type that was found useful first in the context of spin glasses \cite{97FP}.
Following Refs.~\cite{97FP, 10CCGGGPV,12YM,13Ber,14BC,14BCTT,15BJ,15NBC} we can define  an order
parameter with the idea of comparing two different glassy configurations $\{\B r^{(1)}_i\}_{i=1}^N$ and $\{\B
r^{(2)}_i\}_{i=1}^N$ ,
\begin{equation}
Q_{12} \equiv \frac{1}{N}\sum_i^N\theta(a-|\B r^{(1)}_i - \B r^{(2)}_i|)\ ,
\label{Q12}
\end{equation}
wherein $\theta(x)$ is the Heaviside step function. The value of the parameter
$a$ is free, and is determined by trial and error. The quantity $Q_{12}$  is
called an ``overlap" since it has a value that goes from $0$ (completely
decorrelated configurations) to $1$ (perfect correlation). Its purpose is to
measure the degree of similarity between configurations.

Let us now consider a glass, made by quenching a super-cooled liquid with $N$ particles down to a
certain temperature $T\ge 0$ at a suitable rate. A glass is an amorphous solid
wherein particles vibrate around an amorphous structure. So, if we take two
configurations $\{\B r_i^{(1)}\}_{i=1}^N$ and $\{\B r_i^{(2)}\}_{i=1}^N$ from this glass, they will be most
likely close to each other with $Q_{12}$ of the order of unity. If one is able
to obtain a good sampling of the typical configurations visited by the particles
in the glass, one can measure the probability distribution of the overlap
$P(Q_{12})$, which will be strongly peaked around an average value $\langle
Q_{12}\rangle$ close to unity. The configurations visited by the particles will
then form a small connected ``patch" in the configuration space of the system,
selected by the amorphous structure provided by the last configuration that was
visited by the liquid glass former before it fell out of equilibrium while forming
a glass.

Imagine now that we begin to strain this glass. While the stress increases,
there exist plastic events that begin
to cause irreversible displacements in the particle positions. Our order
parameter $Q_{12}$ will begin to respond
to these displacements and will begin to reduce from $O(1)$ to lower values. We
will show now that all along the
``elastic" branch $\langle Q_{12}\rangle$ will remain around unity, but as the
mechanical yield takes place a sharp phase transition occurs, whereupon sub-extensive
plastic events \cite{09LP2,12DHP,13DGMPZ} begin to cause
substantial displacements, allowing different
regions of the configuration space to affect the order parameter.
In such a situation, the distribution $P(Q_{12})$ will have two peaks: one at
high $Q_{12}\le 1$ corresponding to configurations in the same patch and one
for $Q_{12} \ge 0$ corresponding to configurations in different patches.

To demonstrate this fundamental idea we can use any model glass, since this
order parameter description is expected to be universal. For concreteness we
performed molecular dynamics simulations
of a Kob-Andersen 65-35\% Lennard Jones Binary Mixture in $2d$. We have two
system sizes, $N=500$ and $N=4000$. We chose $Q_{12}$ with $a = 0.3$ in LJ
units, but verified that changes in $a$ leave the emerging picture invariant.
As a first step, we prepared a glass by equilibrating the system at $T=0.4$, and
then quenching it (the rate is $10^{-6}$) down to $T=1\cdot10^{-6}$ into a
glassy configuration. The sample is then heated up again to $T=0.2$, and a starting
configuration of particle positions is chosen at this temperature. Note that while at
$T=0.4$ equilibration is sufficiently fast, at $T=0.2$ the computation time is much shorter
than the relaxation time. The
configuration is then assigned a set of velocities randomly drawn from the
Maxwell distribution at $T=0.2$, and these different samples are then quenched down to
$T=0$ at a rate of $0.1$. This procedure can be repeated any number of times (say 500 times),
and it allows us to get a sampling of the configurations inside one single ``patch". We
verify that the typical overlap of the ensemble of inherent structures so
obtained in one patch is close to $\langle Q_{12}\rangle = 1$, signaling that indeed the ensemble
is completely located in a single patch.  Having one patch, we repeat the
procedure starting from another equilibrated configuration of the liquid to create another patch.
The results shown below for the system
with 4000 particles were
obtained by having 100 different patches, each of which containing 500 different inherent structures
due to the velocity randomization. The results with 500 particles were based on 520 different patches
each of which having 100 different inherent structures.
\begin{figure}
\vskip -0.5 cm
\includegraphics[scale = 0.50]{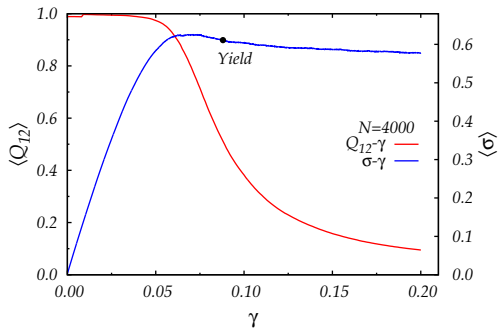}
\caption{a superposition of a stress vs strain curve on  the dependence of $\langle Q_{12}\rangle$ on $\gamma$. The stress vs.
strain curve is obtained by averaging
over 50,000 realizations of individual such curves obtained from 100 initial patches, each of which
containing 500 inherent structures. The order parameter $Q_{12}$ was
averaged on the same realizations to provide $\langle Q_{12}\rangle$.}
\label{FigQ12}
\end{figure}

We then apply to each inherent structure an AQS protocol as described above.
This will create for each value of $\gamma$ a \emph{strained ensemble} of
configurations in the patch whose $P_\gamma(Q_{12})$ is measured. The order parameter is
computed by using {\em all} the unique pairs of configuration generated in the strained
ensemble at a given $\gamma$. We present the
results for $N=4000$ in Fig.~\ref{FigQ12}. We can see how the initial
ensemble for $\gamma = 0$ shows a value of the order parameter $Q_{12} = 1$, signifying
that our initial ensemble is genuinely within one patch. As the ensemble is
strained, the value of the order parameter gets lower, dropping towards zero when the strain
is increased beyond the yield strain.

\begin{figure}
\vskip -0.5 cm
\includegraphics[scale = 0.50]{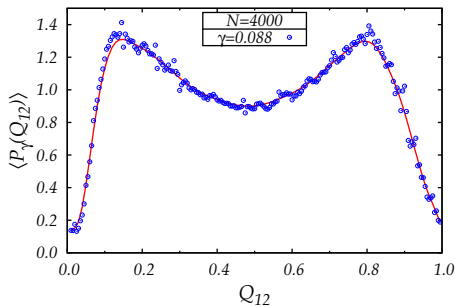}
\caption{The probability distribution function $P_\gamma(Q_{12})$ at $\gamma_{_{Y}}=0.088$ averaged over
100 initial configurations each of which has 500 different realizations to obtain $\langle P_\gamma(Q_{12})\rangle $. At this value of the strain the pdf has two peaks of equal heights. We identify this value of $\gamma$
as the point of the phase transition.}
\label{yield-point}
\end{figure}

To determine the yield strain $\gamma_{_Y}$ accurately, we consider the probability distribution function (pdf)
$P_\gamma(Q_{12})$. We determine $P_\gamma(Q_{12})$ for each patch of of 500 configurations obtained as explained above, and then average the result over the 100 available patches.
We ask at which value of $\gamma$ this averaged pdf has two equally high peaks, see Fig~\ref{yield-point}. The resulting $\langle P_\gamma(Q_{12})\rangle$ determines the yield point to occur at $\gamma_{_{Y}}\simeq 0.088$. Note that this criterion
implies a sharp definition of ``yield" which seems absent in the current literature. If accepted, it indicates that
the mechanical yield occurs beyond the stress overshoot in correspondence with the mean-field results of Ref.~\cite{15RUYZ}.

\begin{figure}
\includegraphics[scale = 0.50]{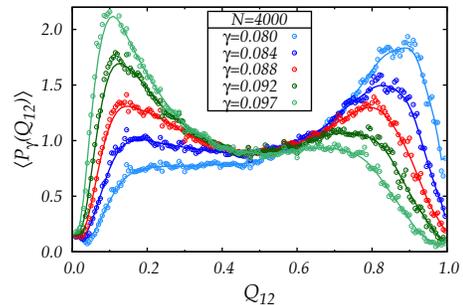}
\caption{The probability distribution function $\langle P_\gamma(Q_{12})\rangle$ in the vicinity
of the critical point $\gamma_{_{Y}}=0.088$}
\label{transition}
\end{figure}

Once we identify the phase transition point, we can demonstrate the transition itself.
In Fig.~\ref{transition} we display the change in $\langle P_\gamma(Q_{12})\rangle$ in the vicinity of the critical
point $\gamma_{_{Y}}$ as a function of $\gamma$.
Within a very narrow range of $\gamma$, of the order of $\Delta \gamma \approx 0.017$, we observe a first-order
like transition from
a pdf with dominant peak at high values of $Q_{12}$ to a dominant peak at low values of $Q_{12}$.   We capture a very
unambiguous and qualitative change in behavior as the yielding point is reached.

\begin{figure}
\includegraphics[scale = 0.50]{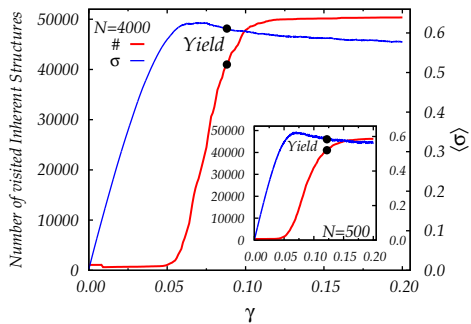}
\caption{The number of configurations which pass below the threshold value $Q_{12}=0.8$ of the overlap order
parameter as a function of
the strain $\gamma$ for $N=4000$. In the onset we show the same test for $N=500$. The conclusion is that {\em all the configurations} lose overlap with the initial
configuration in the vicinity of the yield point $\gamma_{_Y}$}.
\label{number}
\end{figure}

To sharpen the understanding of what is happening in the vicinity of the yield point we examine next {\em how many}
of our realizations loose the tight overlap with the initially prepared configuration and where the loss of
overlap is taking place. To this aim we consider, for the system of 4000 particles, all the 50,000 realizations
that we have. These are obtained by 100 choices of liquid realizations, each of which is velocity randomized 500
times (chosen with Boltzmann probabilities).  When the strain $\gamma$ is increased in our AQS algorithm, we keep
computing the order parameter $Q_{12}$ where the first configuration $\{\B r^{(1)}_i\}_{i=1}^N$ in Eq.~(\ref{Q12}) is
chosen randomly from all the available configurations at that value of $\gamma$,
and the second is any one of the other available configurations at the same value of $\gamma$. We confirmed that
changing the randomly chosen $\{\B r^{(1)}_i\}_{i=1}^N$ does not affect the results. Next, choosing $Q_{12}=0.8$ as a threshold value, we now count how many of our observed configurations cross this threshold and exhibit $Q_{12}\le 0.8$.
The number of configurations that do so as a function of the strain (superimposed on the stress.vs. strain curve) is
shown in Fig~\ref{number}.
The conclusion of this test is that in the vicinity of the yield point $\gamma_{_Y}$ {\em all the configurations}
lose their overlap with the initial configuration, but {\em not before}. The mechanical yield is tantamount to the
opening up of a vast number of possible configurations, whereas before yield the system is still constrained to
reside in the initial meta-basin of the free energy landscape. This appears to be a first-order phase transition \cite{16RU}.

\begin{figure}
\includegraphics[scale = 0.40]{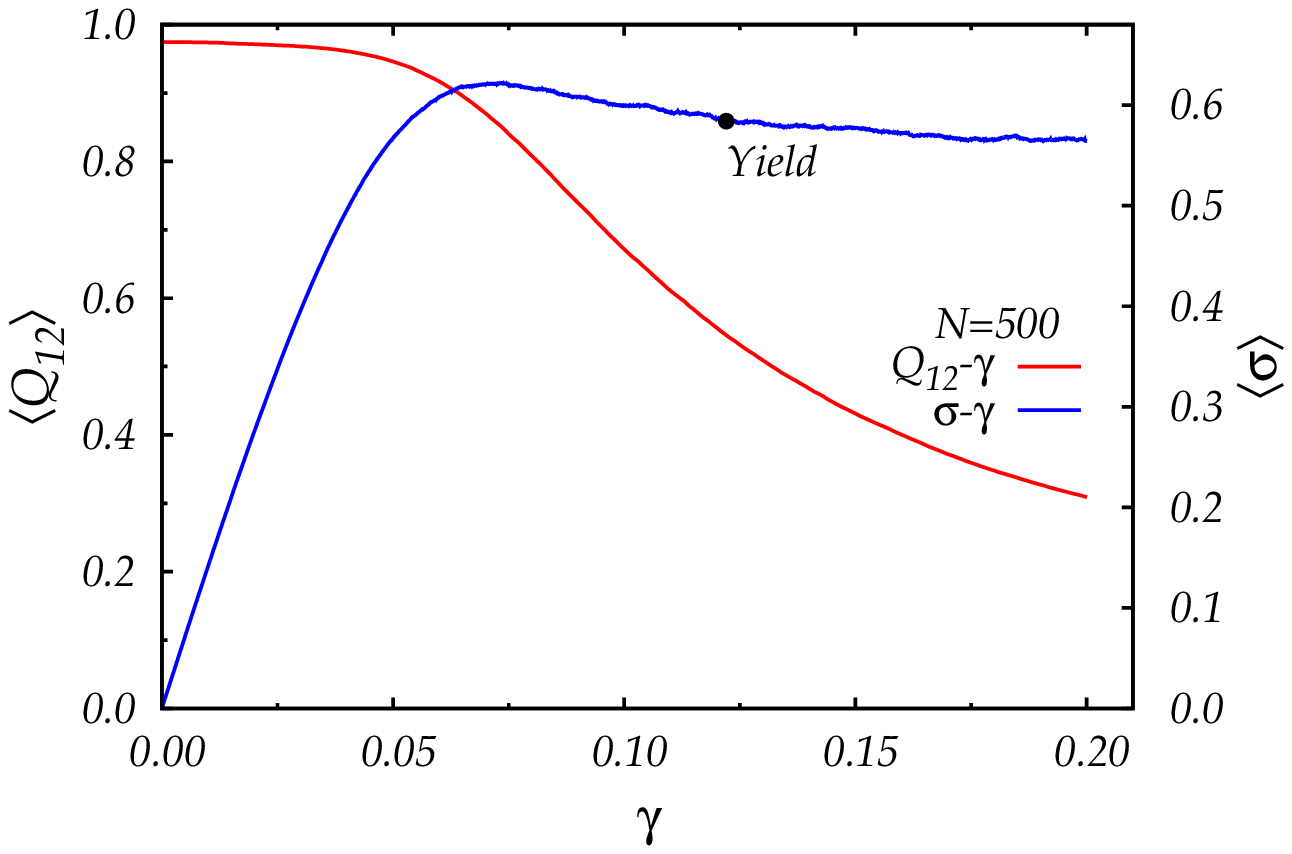}
\includegraphics[scale = 0.40]{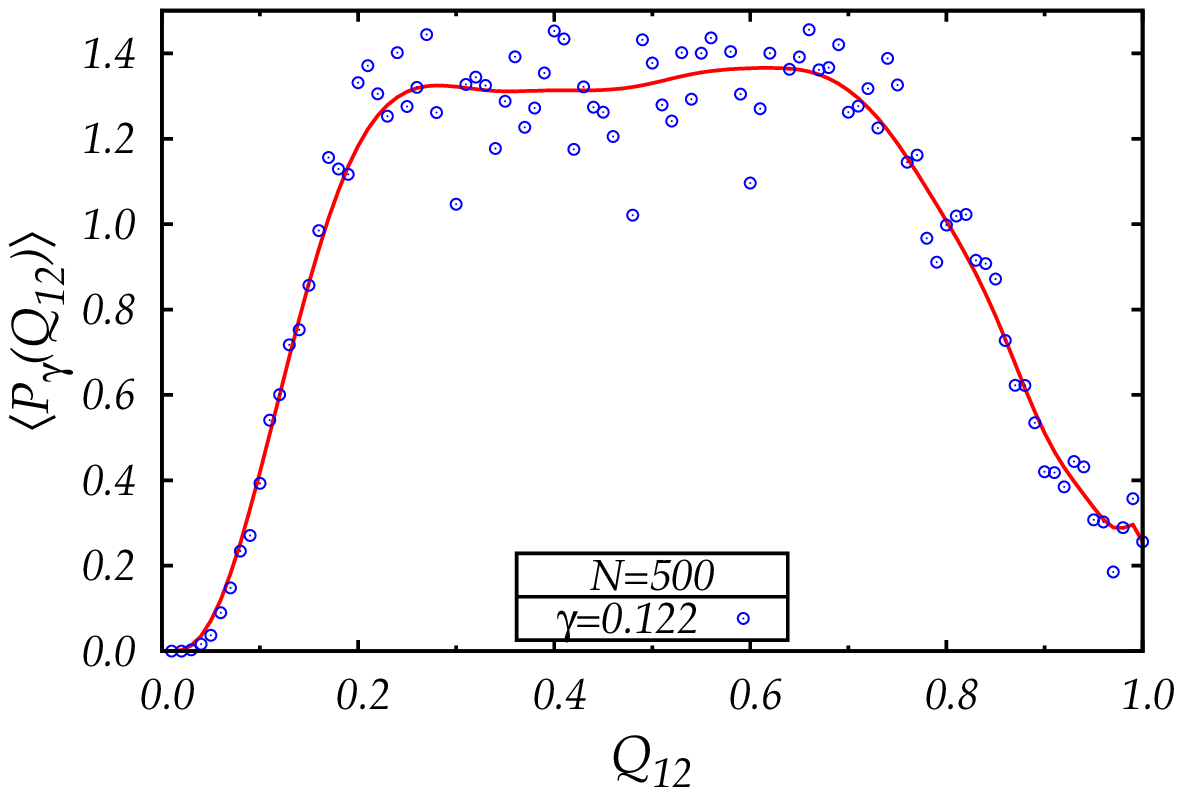}
\includegraphics[scale = 0.40]{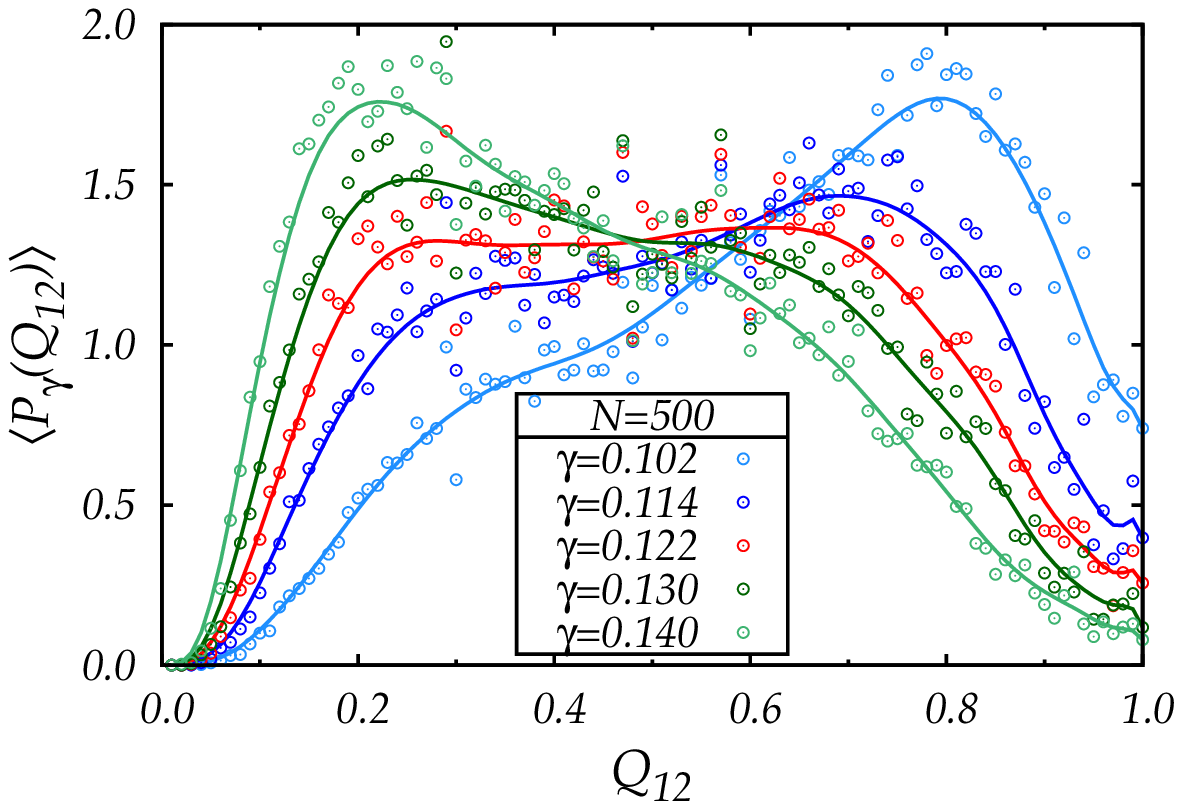}
\caption{Results similar to those shown in Figs.2-4 but for system size
$N=500$. One observes a smearing out of the transition region as is expected from a first-order phase transition.}
\label{N500}
\end{figure}

To strengthen the proposition that this is a first-order phase transition we should demonstrate
that the transition becomes sharper with increasing the system size. At present we cannot produce equally good data
for systems of size much larger than $N=4000$, but we have produced equally extensive data for $N=500$.
The results for this smaller system size are presented in Fig.~\ref{N500}.
Indeed, the change in the values of $Q_{12}$ diminishes as seen in the upper panel, the identification of the transition
point is less sharp, and most importantly, the range of $\Delta \gamma$ over which a similar change in the peak structure
is taking place is now $\Delta\gamma\approx 0.038$. If we take just these two system sizes as indicative we can roughly
estimate the range of $\Delta \gamma$ over which the transition is taking place to go to zero as $N^{-1/3}$ as $N\to \infty$.
Needless to say, at this point this should be taken as indicative only, and further accurate simulations should be conducted
to solidify (excuse the pun) this important issue.

The upshot of these results is that we are able to put a finger on
the essential feature that
is responsible for the mechanical yield. It is not that the configurations
visited by the system after yield have
different characteristics from the configurations before yield. Rather, a
very constrained set of configurations
available to the system before yield is replaced upon yield with a vastly larger
set of available configurations. This much larger set is generic; it is not selected by any careful cooling protocol,
and as such it is expected to include many marginally unstable configurations that will yield plastically with any
increase of strain \cite{10KLP,16HJPS}. This is the fundamental reason for the inability of the system to continue to
increase its stress when strain is increased, leading to the steady state branch. We propose this as a universal
mechanism for the ubiquitous prevalence of stress vs strain curves that look so similar in a huge variety
of glassy systems.

\acknowledgments
This work had been supported in part by an ``ideas" grant STANPAS of the ERC and by the Minerva
Foundation, Munich, Germany. We thank Ludovic Berthier for clarifying the role of ``overlap" order
parameters in distinguishing amorphous configurations. Discussions with Oleg Gendelman and Carmel Shor
are gratefully acknowledged.


\end{document}